\documentclass[conference]{IEEEtran}
\IEEEoverridecommandlockouts
\usepackage{cite}
\usepackage{amsmath,amssymb,amsfonts}
\usepackage{algorithmic}
\usepackage{graphicx}
\usepackage{textcomp}
\usepackage{xcolor}
\usepackage{cleveref} 
\usepackage{enumitem} 
\usepackage{url} 
\Crefname{figure}{Fig.}{Figs.}
\def\BibTeX{{\rm B\kern-.05em{\sc i\kern-.025em b}\kern-.08em
    T\kern-.1667em\lower.7ex\hbox{E}\kern-.125emX}}

\makeatletter
\newcommand{\linebreakand}{%
  \end{@IEEEauthorhalign}
  \hfill\mbox{}\par
  \mbox{}\hfill\begin{@IEEEauthorhalign}
}
\makeatother

\begin{document}

\title{Enterprise-Driven Open Source Software: \\ A Case Study on Security Automation}

\author{\centering \IEEEauthorblockN{Florian Angermeir}
\IEEEauthorblockA{Technical University of Munich\\
Germany \\
https://orcid.org/0000-0001-7903-8236}
\and
\IEEEauthorblockN{Markus Voggenreiter}
\IEEEauthorblockA{Siemens Technology and Ludwig Maximilians University Munich \\
Germany \\
https://orcid.org/0000-0003-3964-1983}
\linebreakand
\IEEEauthorblockN{Fabiola Moyón}
\IEEEauthorblockA{Siemens Technology and Technical University of Munich \\
Germany\\
https://orcid.org/0000-0003-0535-1371}
\and
\IEEEauthorblockN{Daniel Mendez}
\IEEEauthorblockA{Blekinge Institute of Technology and fortiss GmbH\\
Sweden and Germany \\
https://orcid.org/0000-0003-0619-6027}
}

\maketitle

\begin{abstract}
    Agile and DevOps are widely adopted by the industry. Hence, integrating security activities with industrial practices, such as continuous integration (CI) pipelines, is necessary to detect security flaws and adhere to regulators' demands early.
In this paper, we analyze automated security activities in CI pipelines of enterprise-driven open source software (OSS). This shall allow us, in the long-run, to better understand the extent to which security activities are (or should be) part of automated pipelines. 
In particular, we mine publicly available OSS repositories and survey a sample of project maintainers to better understand the role that security activities and their related tools play in their CI pipelines. To increase transparency and allow other researchers to replicate our study (and to take different perspectives), we further disclose our research artefacts. \\
\noindent Our results indicate that security activities in enterprise-driven OSS projects are scarce and protection coverage is rather low. Only 6.83\% of the analyzed 8,243 projects apply security automation in their CI pipelines, even though maintainers consider security to be rather important. This alerts industry to keep the focus on vulnerabilities of 3rd Party software and it opens space for other improvements of practice which we outline in this manuscript.

\end{abstract}


\section{Introduction}
\noindent The rise of agile software development and DevOps contributed to shortening the software development life cycle while aiming at increasing efficiency and customer value \cite{turpe2017managing}. This is mainly achieved by faster development cycles and feedback loops where the software development process is highly equivalent to the production process in a factory \cite{humble2010continuous,kim:2016:devops}. The output of one phase is input to the next one. The ability to recognize failed outputs or to keep up continuity in the production process became a key aspect to maximize development throughput (see, e.g., the fail early approach and the pipeline concepts).

\noindent Agile and DevOps are both becoming the state-of-practice for continuous software development in several industries, such as ours. Due to their wide adoption, integrating security aspects into continuous processes has become a must to improve software practice \cite{Othmane:2014,Fitzgerald:2017,morales:2020:guide}. 

\noindent However, security needs to be systematically integrated to ensure that the benefits to continuous processes remain. For example, agile and DevOps build upon the mindset that teams are the owners of the software process lane, thus, security integration should not only focus on protecting the product, but also on enabling teams to take ownership for security to ensure protection in the long run \cite{Rahman:2016}. Specifically in DevOps, the factory analogy for software development sets automation -- an exclusively technical aspect -- as the main goal \cite{Soni:2015,Jabbari:2016}. For security purposes, automation could be even \textit{the panacea}. Which security responsible or incident manager would not dream about automating each single security test so that developers solve issues in time, directly along ``their'' process lane, same as in a factory? Although current practice offers several possibilities, security integration should not only consider proper tools but the current involvement of them in the pipeline. This goes beyond the plug-in approach by additionally considering people and process aspects. An example is how teams choose appropriate security tools and how they clarify whether security tools need extra processes that ensure issues are solved, documented, or escalated.

\noindent In this work, we approach one of the several issues for agile and DevOps: security automation in continuous integration (CI) pipelines. In our experience from Finance, Health, and Critical Infrastructure industries, development teams are urged by regulators to involve security activities in their development processes\cite{moyon2020compliantdevops,yasar:2017:implementing}. This implies that CI pipelines should include checks to identify a set of security issues. In order to implement these checks, additional security tools are required. 

\noindent Challenges arise when industry software products involve external components e.g. open source software (OSS). First, industrial companies need to ensure that external component providers involve security activities in the development processes. Second, development teams, both industrial and external, should select security tools that accurately automate such activities. Security tools in CI pipelines become information assets to protect, as well as enablers for new attack vectors.

\noindent To this end, we analyze which security activities are involved in CI of enterprise-driven OSS projects, meaning projects driven or at least mainly contributed to by enterprise employees. Our study allows practitioners to determine the coverage of security activities in external products while depicting the most used security tools. Software developers may use our results to evaluate possible tools for their industrial environments.

\noindent In detail, we investigate security tools in CI pipelines of publicly available OSS projects. Based on the tools, we infer security activities. We further validate our findings by approaching a sample of the projects' maintainers, to understand the role security plays in their projects and how project characteristics correlate with the security.

\noindent Our analysis is based on the dataset published by Spinellis et al.\cite{Spinellis:2020}, consisting of enterprise-driven open source software using GitHub as code repository. With our approach, we address one of the future research ideas of Spinellis et al. with a focus on security tools usage. By mining this dataset, we aim to address the following issues: Considering solely enterprise-driven projects addresses the generalizability concerns of analyzing open source software projects. Thus rendering our outcomes applicable to our industrial setting. Furthermore, by analyzing the dedicated data set comprising data from various enterprises we aim at tackling the problem of data gathering. Conducting a survey-based study among several enterprises introduces the challenge of a potentially limited sample size. However, conducting it solely among one enterprise, induces relevance concerns. Consequently, employing the given dataset overcomes both challenges. \\
As framework for our study, we aim to answer the following research questions:
  \begin{itemize}
  \setlength\itemsep{2pt}
            \item \textbf{RQ1:} How prevalent are security activities in continuous integration of enterprise-driven open source software?
            \item \textbf{RQ2:} Which types of security tools are utilized in CI of enterprise-driven open source software?
            \item \textbf{RQ3:} Do certain types of projects perform security activities more often than others?
        \end{itemize}

\noindent The rest of the paper is organized as follows: Section~\ref{sec:background_rel_work} summarizes related work about security automation in CI pipelines, as well as, repository mining to identify practices in software engineering. The study methodology is described in Section~\ref{sec:emp_stud}, while  Section~\ref{sec:study_res} contains the results. In Section~\ref{sec:discuss}, we discuss the implications of our results for industrial companies using OSS software and also compare results with our observations in industrial CI. Finally, we conclude in Section~\ref{sec:conclusion}, introducing our ideas for future work.

\section{Background and Related Work}
    \label{sec:background_rel_work}

\subsection{Security Automation}
\noindent Since automation is key for DevOps and agile processes \cite{Leite:2019}, automating security is the next logical step to overcome the challenges of closing the gap between software engineering and compliance with security standards. Existing research describes first ways on introducing security into agile processes and DevOps \cite{yasar:2016:where, Rahman:2016}. Various papers already researched individual security tasks in software engineering. For example, Cadariu et al. \cite{7081868} identified the need for automated known vulnerability detection, in particular with respect to external components and systems. In \cite{10.1145/1501434.1501486} Mourad et al. identify security hardening as a possible concept to help developers and maintainers eliminate and prevent vulnerabilities and threats. Complementing the concept of Mourad et al. the need for automation of security compliance checks was described by Ullah et al. in \cite{6681020}. Through automated checking, whether software complies with predefined security rules, the effort in terms of cost and time can be reduced, while security throughout the whole development cycle is increased. However, even though there exists research on individual security activities, there is only little research on systematically integrating security \cite{moyon2020compliantdevops}. For that purpose, classification of the security tasks has to be performed first. By relying on security standards such as IEC 62443-4-1 \cite{iec4_1} and existing research to categorize security tasks (from now on called security activities) we bridge the gap between security regulators and software engineering practitioners to facilitate the integration of automated security. For the purposes of this paper, the categories of security activities are:

\begin{itemize}
 \setlength\itemsep{3pt}
                \item \textbf{SA1: 3rd Party Vulnerability Scanning} is the verification of correct implementation and testing against known security risks and vulnerabilities in external components.
                \item \textbf{SA2: Static/Dynamic Application Security Testing} refers to testing for security vulnerabilities of product components and their dependencies to maintain an up-to-date security patch level.
                \item \textbf{SA3: Secure Configuration/Hardening} describes the process of applying security configuration measures to minimize the attack surface or even ensure security standard compliance.
                  
                \item \textbf{SA4: Compliance/Hardening Checks} verify that security configuration/hardening measures are aligned with best practices.
                \item \textbf{SA5: Secrets Management} contains the management of sensitive information such as passwords or private keys as well as protective measures to prevent leakage of secret information.
\end{itemize}

\subsection{Mining Repositories}
\noindent Analyzing CI pipelines of open source projects is no new concept. For instance, Widder et al. \cite{8595199} analyzed projects to identify indicators for the reduced usage of Travis CI. Their data set consisted of 1,819 open source software projects on GitHub, which utilized Travis CI but moved to another continuous integration software. Other studies go beyond the continuous integration tooling and focus on the activities and tools used in the CI pipeline. \\
Zampetti et al. \cite{static_code_analysis} conducted a study, which focused specifically on the usage of static code analysis tools in CI pipelines. They analyzed the CI pipelines of 20 Java-based open source software projects. Their study analyses static code analysis tools along with their configuration in pipelines. The limitations of this study are within the limited sample number of 20 projects and the focus on static code analysis as only security activity. Especially the automation of security activities was covered by Yasar in 2018 \cite{8543411}. In their paper, the challenges around secrets management, when automating security activities are addressed. Even though the author identifies the need for automation in CI, their experiment showed a "concerning number of actual cases of password hardcoding".

\noindent Analyzing available data from code repositories like GitHub comes with the risk of working on non-representative data. Consequently, the data source for this paper consists of the database by Spinellis et al. \cite{Spinellis:2020}. They elucidate, that working on data from enterprise-driven software development projects addresses generalization issues that come with volunteer developed projects. In order to create the dataset, they filtered a list of existing GitHub projects. As the premise for the filtering, they relied on contributors using their corporate email when interacting with GitHub projects. The resulting 17,264 GitHub projects are used as data foundation for this paper.

\section{Empirical Study Design}
    \label{sec:emp_stud}
\begin{figure*}
        \centerline{\includegraphics[width=\textwidth]{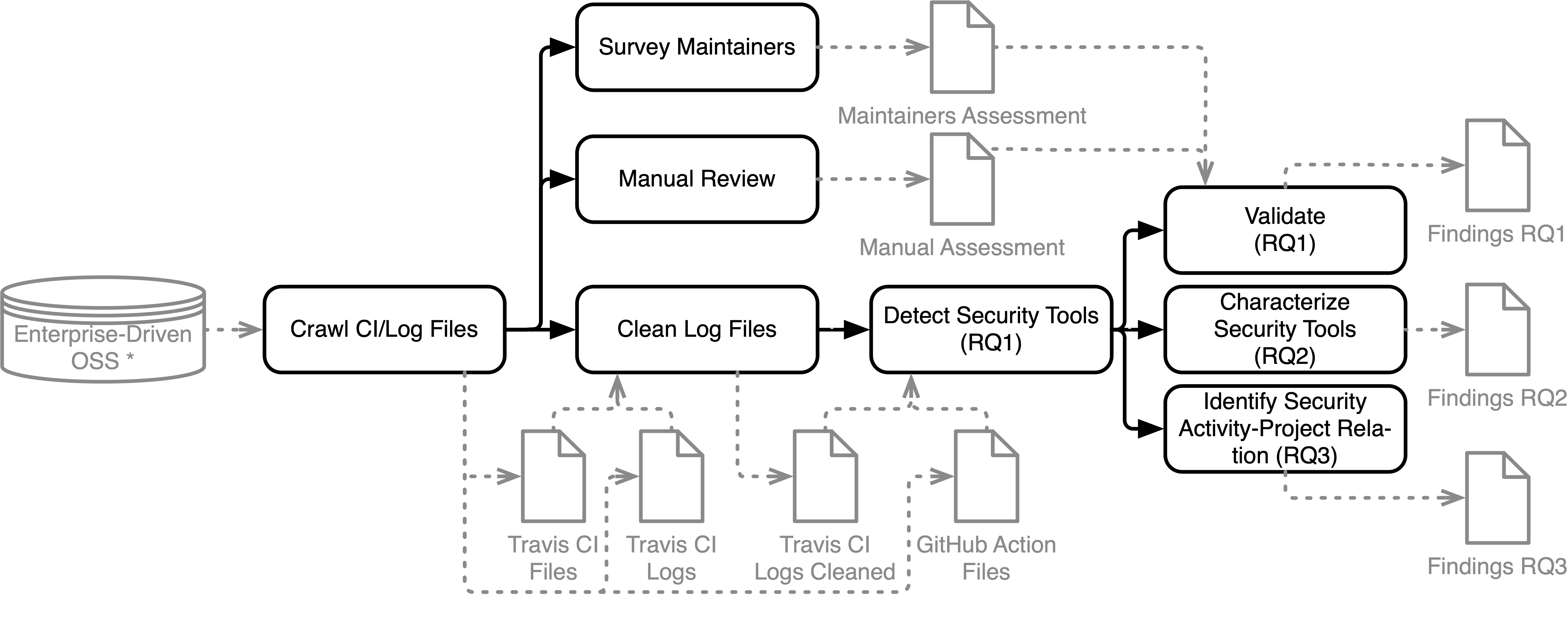}}
        \caption{Detailed overview of the applied methodology in this study. *based on \cite{Spinellis_dataset:2020}}
        \label{fig:methodology}
    \end{figure*}

    \noindent This study is composed of several parts. \Cref{fig:methodology} depicts a detailed overview of the methodology applied. First, we investigated the prevalence of security activities in CI of enterprise-driven OSS by analyzing open source software projects on security tool usage. Second, we validated these findings through manual review by the authors and surveys with the project's maintainers. Based on previously collected data we investigated the types of used security (relevant) tools. Finally, we explored whether certain project characteristics relate to security tool usage. \\
    The artefacts of our study can be consulted in our online material at: \url{https://doi.org/10.5281/zenodo.4106329} \cite{we:2020}.
    
    \subsection{Research Questions}
    \noindent The study of this paper addresses the following  research questions:
        \begin{itemize}
         \setlength\itemsep{2pt}
            \item \textbf{RQ1:} \textit{How prevalent are security activities in continuous integration of enterprise-driven open source software?}
            \begin{itemize}[leftmargin=*]
                \item How many projects perform security activities?
                \item What is the distribution of security activities?
                \item Which security activities are performed together the most?
                \item What are reasons for not performing security activities?
            \end{itemize}
            \item \textbf{RQ2:} \textit{Which types of security tools are utilized in CI of enterprise-driven open source software?} \\
            Based on the results of RQ1 we determine security-relevant tools used in CI of enterprise-driven OSS and their respective usage frequency. We then investigate possible causes for their usage such as cost or integration effort.
            \item \textbf{RQ3:} \textit{Do certain types of projects perform security activities more often than others?} \\
            We examine whether certain project characteristics like popularity or used languages relate to the usage of particular security activities.
        \end{itemize}
        \label{RQs}
        
    \subsection{Data Collection}
        \label{sec:datacol}
	    \noindent We based the data collection of this study on the enterprise-driven open source software dataset, presented by Spinelli et al. at the Mining Software Repositories Conference 2020 \cite{Spinellis_dataset:2020}. This dataset contains 17,264 GitHub projects driven, or at least mainly contributed to, by enterprise employees. The data collection process contains manual as well as automatic steps, which are explained in the following.
	    
	 \subsubsection{Automatic Collection Steps}
	    Out of 17,264 GitHub projects, 16,742 were still available at collection time (13.08.2020-16.08.2020). To lay the foundation for the answers of RQ1 and RQ2 we crawled each project's code repository for files of four CI services: Travis CI (.travis.yml), GitHub Actions (.github/workflow/), GitLab CI (.gitlab-ci.yml), and Circle CI (.cirleci/). These services were selected since they are widely used and mainly configured through a CI file stored in the repository with a fixed naming scheme, which enabled us to detect their usage and further investigate them. In total 8,954 projects used at least one of these continuous integration services. As depicted in \Cref{table:CI_service_usage}, the most used services were Travis CI and GitHub Actions with a total usage of over 100\%. This is due to the fact, that 1,513 (16.90\%) of these projects used both services. Based on the coverage of Travis CI and GitHub Actions we decided to focus on these two, which left 8,423 projects for the analysis. Additionally to the CI files stored in the repositories, we crawled the CI logs of the last three pipeline executions for a more effective security tool usage detection since not all tools are configured through a CI file. However, we could only crawl logs for Travis CI since GitHub Action logs' availability is limited to 90 days. This restriction made it infeasible to crawl a reasonable amount of GitHub Action logs \cite{github:2020}. To answer RQ3 we additionally crawled certain characteristics of each project. These characteristics are the number of stars, the programming languages of the project and the year it was created, as well as the year it was last updated.
        
        \begin{table}[b]
            \caption{Distribution of Continuous Integration Services Usage}
            \begin{center}
                \begin{tabular}{c|c|c|c|c}
                    \hline
                    \textbf{CI Services} & \textbf{\textit{Travis CI}}& \textbf{\textit{GitHub Actions}}& \textbf{\textit{Circle CI}} & \textbf{\textit{GitLab CI}} \\
                    \hline
                        \textbf{Projects} & 6093 & 3843 & 820 & 247 \\
                        \textbf{Distr. in \%} & 68.05\% & 43.92\% & 9.15\% & 2.76\% \\
                    \hline
                \end{tabular}
                \label{table:CI_service_usage}
            \end{center}
        \end{table}
    
    \subsubsection{Manual Collection Steps} 
	    In the manual data collection, we created a list of security tools or security-related tools that fit our definition described in \Cref{sec:background_rel_work}. This list is required for an automated search for security tool usage during the data analysis phase. In total, we selected 84 security tools or tools related to security activities. The distribution is listed in \Cref{table:tool_per_act} and the complete list of tools is available in the online material at \cite{we:2020}. Some of the tools (e.g. Snyk) could be mapped to multiple security activities. To address this concern we decided to only select the primary security activity, which means the activity that is the primary/original focus of the tool. This minimizes the intersection between security activities. Although, primary focus of the tools selected for Secure Configuration/Hardening is not security, they are all related to security. Considering this study's focus is on tool usage detection, we included them as well. Identification of whether these tools actually improve security properties in each individual case is left for future research.
	    \begin{table}[b]
            \caption{Tools per Security Activity}
            \begin{center}
                \begin{tabular}{l|c}
                    \hline
                    \textbf{Security Activities} & \textbf{\#Tools} \\
                    \hline
                        \textbf{SA1: 3rd Party Vulnerability Scanning} & 15 \\
                        \textbf{SA2: Static/Dynamic Application Security Testing} & 25 \\
                        \textbf{SA3: Secure Configuration/Hardening} & 7 \\
                        \textbf{SA4: Compliance/Hardening Checks} & 16 \\
                        \textbf{SA5: Secrets Management} & 21 \\
                    \hline
                \end{tabular}
                \label{table:tool_per_act}
            \end{center}
        \end{table}
	    
		\noindent In the next step, we created a list of patterns to clean the data from package manager and installation instructions prior to the analysis. For the validation, we first manually reviewed 400 randomly select projects for security tools usage. The manual review was performed by going through each CI file and each log file of the project in question. For each tool used or logged, we analyzed whether it can be mapped to a security activities defined in \Cref{sec:background_rel_work}. In addition, we asked project maintainers for their assessment of security tool usage in the respective project via email. To this end, we sent out 664 emails containing the questions stated in  \Cref{table:interview_questions} and examples for the defined security activities as stated in \Cref{table:interview_examples}. Two of the authors individually interpreted the answers and later discussed their interpretation to find a common understanding of the maintainers' assessment. 
        \begin{table}[b]
            \caption{Survey Questions}
            \begin{center}
                \begin{tabular}{c l}
                    \hline
                        \textbf{1.} & Do you perform security activities on this project? \\
                        \textbf{2.} & If yes, which tools do you use? \\
                        \textbf{3.} & \begin{tabular}[c]{@{}l@{}}If no, do you plan to add security activities to this project \\ in the future? If so, which? \end{tabular}\\
                        \textbf{4.} & \begin{tabular}[c]{@{}l@{}}From 1 (not important) to 5 (very important), how important do \\ you perceive security in this project? \end{tabular} \\
                    \hline
                \end{tabular}
                \label{table:interview_questions}
            \end{center}
        \end{table}
        
        \begin{table}[htbp]
            \caption{Security Activity Examples in Survey}
            \begin{center}
                \begin{tabular}{c | l}
                    \hline
                    \textbf{Security Activity} & Tools \\
                    \hline
                        \textbf{SA1} & OWASP Dependency Tracker, WhiteSource, Clair \\
                        \textbf{SA2} & Bandit, Checkstyle, SonarQube \\
                        \textbf{SA3} & Ansible, Chef, Turbot, SecComp \\
                        \textbf{SA4} & Terrascan, Chef inSpec, Docker Bench \\
                        \textbf{SA5} & Vault, TruffleHog, Gitleaks \\
                    \hline
                \end{tabular}
                \label{table:interview_examples}
            \end{center}
        \end{table}

    \subsection{Data Analysis}
      \noindent To start with the data analysis, the log files required cleaning. This was necessary since the output of a package manager instruction like \emph{installation} or \emph{update} can contain security tool patterns, even though this does not imply that the security tool is used. The process for cleaning the log files is as follows:
        \begin{itemize}
            \item[1.] Detect usage of package manager or installation instruction in CI file (instruction 1)
            \item[2.] Detect next instruction in CI file, not associated to the package manager/installation instruction (instruction 2)
            \item[3.] Remove the logged part between instruction 1 (inclusive) and instruction 2 (exclusive)
        \end{itemize}
        To answer RQ1 we defined at least one search pattern for each tool. During the automatic security tool detection, we queried all CI and log files for the predefined patterns. 
        However, some special cases had to be considered during the search operation. Some environment variables can hint at tool usage. As those are often written in uppercase, the search had to be case insensitive. Additionally, a differentiation between tools with a common prefix like ``puppet'' and ``puppet-lint'' was necessary for the search function. Consequently, we suffixed the pattern of the shorter one with a white space so both tools were detected separately. The validation is composed of two stages. First, we compared the results of manual review and automatic tool detection to determine false positives and negatives. Furthermore, this allowed us to assess the effectiveness of our automated detection approach. Second, we compared the results of the automatic tool detection to the project's maintainers assessment and identified causes for false negatives of our approach. Using the answers we further investigated RQ1. To determine which types of security tools are utilized we initially identified the tools actively used by projects in the dataset. Afterward, we investigated their properties, like costs, integration effort, or use case. Taking care of RQ3, we analyzed the relation between the project's characteristics and security tool usage. For that purpose, we compared the characteristics of projects performing security activities to the characteristics of those not performing security activities. Here, we first focused on the number of stars per project by calculating the lower quartile, the median, and the upper quartile. Following this approach, we analyzed the year the project was created, last updated and for how long it has been maintained, also measured in years. Finally, we compared the programming languages used by projects performing different security activities. For each project, the \emph{language usage} was measured in bytes of code programmed in that language. We normalized the language usage by dividing by the total number of code bytes comprised by the project. For each language, we summarized the normalized language usage of all projects and mapped the security activities belonging to the separate projects. This results in a mapping of security activities to programming languages. 
\section{Empirical Study Results}
    \label{sec:study_res}
    \noindent In this section we describe the results from the empirical study for each research question and validate them accordingly.
    
    \subsection{RQ1: How prevalent are security activities in CI of enterprise-driven OSS?}
        \textit{How many projects perform security activities?} \\
        \noindent In total, we detected 575 projects performing security activities out of 8,423 analyzed projects after cleaning the data. The cleaning process reduced the number of false positives by 765. That means 6.83\% of all projects using continuous integration (CI) seem to perform some type of security activity.
        \begin{center}
            \fbox{\parbox[c]{0.9\columnwidth}{Around 6.83\% of the analyzed projects perform security activities in continuous integration.}}
        \end{center} \vspace{0.2cm}
        
        \textit{What is the distribution of security activities?} \\
        \noindent \Cref{table:sec_act_distr} depicts how many projects perform a certain security activity. The two top activities are 3rd Party Vulnerability Scanning (261 projects) and Static/Dynamic Application Security Testing (224 projects). Even though we scanned for fewer tools related to Secure Configuration/Hardening than Compliance/Hardening Checks and Secrets Management (see \Cref{table:tool_per_act}, it was performed more often (114 projects) than the later two (50 and 40 projects). We will see indications for the low usage of Compliance/Hardening Checks in the validation phase.
        \begin{center}
            \fbox{\parbox[c]{0.9\columnwidth}{3rd Party Vulnerability Scanning and Static/Dynamic Application Security Testing are performed the most.}}
        \end{center} \vspace{0.2cm}

        \textit{Which security activities are performed together the most?} \\
        \noindent While 491 projects performed one security activity, 83 projects performed multiple. The security activities performed together the most are 3rd Party Vulnerability Scanning and Static/Dynamic Application Security Testing with 26 projects. This selection of tools is reasonable from a security perspective as it covers vulnerabilities induced by 3rd parties as well as ones caused by own code.  This is followed by Secure Configuration/Hardening and Compliance/Hardening Checks with 24 projects which makes sense, since these two complement each other. Any other combination is performed by less than 10 projects, thus we omit them at this point.
        \begin{center}
            \fbox{\parbox[c]{0.9\columnwidth}{3rd Party Vulnerability Scanning and Static/Dynamic Application Security Testing are performed together the most, followed by Secure Configuration/Hardening and Compliance/Hardening Checks.}}
        \end{center} \vspace{0.2cm}
        \begin{table}[b]
            \caption{Distribution of Performed Security Activities}
            \begin{center}
                \begin{tabular}{l|c}
                    \hline
                    \textbf{Security Activity} & \textbf{\textit{Number of Projects}} \\
                    \hline
                        \textbf{SA1: 3rd Party Vulnerability Scanning} & 261  \\
                        \textbf{SA2: Static/Dynamic Application Sec. Testing} & 224\\
                        \textbf{SA3: Secure Configuration/Hardening} & 114  \\
                        \textbf{SA4: Compliance/Hardening Checks} & 50  \\
                        \textbf{SA5: Secrets Management} & 40  \\
                    \hline
                \end{tabular}
                \label{table:sec_act_distr}
            \end{center}
        \end{table}
        
        \textit{Validation} \\
        \noindent In the validation phase, we first manually reviewed 400 projects for security tool usage. In direct comparison to the results of automatic detection, we identified eight projects for that the automatic detection yielded false positives. This is due to two reasons. First, the log file cleaning missed package manager output when called in a conditional instruction, and second, sometimes tool names also appear in other contexts like the tool \textit{phan} in the email address \textit{stephan@example.com}. In addition to the 8 false positives, we detected 31 false negatives, meaning tool usage that was not automatically detected. The activity distribution of false positives is 11 projects for SA1, 17 for SA2, and 1 for SA3, SA4, and SA5 respectively. Besides validation by the authors, we also asked the maintainers for their assessment on security tool usage in the respective project. From 664 emails we sent out, 67 maintainers answered. Of these 67 projects, our detection results differed from the maintainers' answers in 25 projects. This includes false positives, false negatives as well as partial false positives and negatives. We define three categories for the differences: 
        \begin{enumerate}
            \item Tools we didn't search for
            \item Tools we searched for but didn't detect
            \item Tools suggested by the users, but out of scope
        \end{enumerate}
        The before mentioned 25 projects can be classified as follows: 7 projects used tools we didn't search for like Rubocop, Klocworks, or CredScan, 17 projects used tools we searched for, but could not detect. The main reason for this result is the limitation of the detector. While many tools are configured in CI files, some tools like Dependabot are configured outside. Thus the only way to detect these tools is to analyze additional configuration files in the repository or CI logs, which we, as explained in \Cref{sec:datacol}, did not look into for GitHub Actions. Various maintainers answered with a range of tools that are out of scope for this study. Examples are tools related to functional testing such as Google's Sanitizer or related to formatting such as scalafmt. \\
        
        \textit{What are reasons for not performing security activities?} \\
        \noindent From the answers we received, 37 projects did not perform security activities at all, even though they rated the importance of security in their project as 2.1 on average (on a scale from 1 to 5). For comparison, the maintainers that perform security activities rated the importance as 3.7 on average. Maintainers stated various reasons for not performing security activities. Most often the maintainers see the responsibility for security on user/integrator side: \vspace{0.1cm}
        \begin{itemize}[leftmargin=*]
            \item[] \textit{``The final integrators of this project/library are responsible for the security of the product they deliver to customers.''}
            \item[] \textit{``We expect the users to have sufficiently secure and controlled environment.''}
        \end{itemize} \vspace{0.1cm}
        This could also explain why so few projects perform Secure Configuration/Hardening and Compliance/Hardening Checks. Others estimate the attack surface of their project as too limited, either because of its design or the intended usage:
        \begin{itemize}[leftmargin=*]\vspace{0.1cm}
            \item[] \textit{``[...] thus although security remains critical, we have intentionally designed this application to minimize risk in this area.''}
            \item[] \textit{``This is a GUI application that the user runs locally (maybe even sandboxed in snap).''}
        \end{itemize} \vspace{0.1cm}
        Some maintainers perform security activities manually through review by internal departments before integration: \vspace{0.1cm}
        \begin{itemize}[leftmargin=*]
            \item[] \textit{``So you could say we secure at integration, not at development.''}
            \item[] \textit{``We do not regularly do so [perform security activities] any longer but we do on occasion.''}
        \end{itemize} \vspace{0.1cm}
        Five maintainers plan to add or would be open to introduce security activities to their project, either by automating existing manual security activities or by introducing new tools. Those that would be open to introduce security activities lack enough knowledge about existing security tools: \vspace{0.1cm}
        \begin{itemize}[leftmargin=*]
            \item[] \textit{``If there were tools available, I would certainly run them as part of the CI build process.''}
            \item[] \textit{``I would add automated security scanning to the Docker image if it were easy and free.''}
        \end{itemize} \vspace{0.1cm}
        Out of all responses three maintainers couldn't give information about their security activities since it is confidentially:
        \begin{itemize}[leftmargin=*]
            \item[] \textit{``We do not disclose our security strategy and procedures.''}
            \item[] \textit{``[...] security is a touchy subject, and it's not something I'll give details out to strangers.''}
        \end{itemize} \vspace{0.1cm}
        \begin{center}
            \fbox{\parbox[c]{0.9\columnwidth}{Reasons for not performing automated security are:
            \begin{itemize}
                \item user's responsibility to ensure security
                \item attack surface too limited
                \item manual review before release
            \end{itemize}}}
        \end{center} \vspace{0.1cm}
    \subsection{RQ2: Which types of security tools are utilized in CI of enterprise-driven OSS?}
        \noindent Dependabot\footnote{\url{https://dependabot.com/}} is by far the most used tool with 169 projects. It is a 3rd Party Vulnerability Scanning Tool supporting 15 programming languages, deeply integrated into GitHub and for free use. It is followed by CodeQL\footnote{\url{https://codeql.com}}, a static code analyzer supporting eight programming languages. It is also deeply integrated into GitHub and used by 68 projects from the dataset. The low integration effort for both tools also explains the high usage numbers of SA1 and SA2 from RQ1. The list is then followed by Chef (49 projects), Checkstyle (44 projects), Ansible (36 projects) and Coverity (33 projects). All detected tools are listed in \Cref{table:sec_tool_distr} along with the number of projects using it and the respective security activity.
        41 of the 51 tools are free or partially free of charge. So clearly more free than commercial tools are used, however, both commercial (e.g. Snyk, Blackduck, Checkmarx) and free tools (e.g. Dependabot, InSpec, Trufflehog) are distributed across all usage numbers. Also, popularity does not seem to be an accurate indicator. Even though most of the tools among the top 15 are stared more than 1,000 times on GitHub, there are also tools like Pinterest Knox or Bane stared 865 and 838 times. The same applies to the opposite. So the least used tools also contain some tools stared with more than 1,000 stars like Trufflehog, Qark, and conftest. Also, the diversity within a security activity plays a role. While only seven tools are used for Secrets Management, the diversity of tools in Static/Dynamic Application Security Testing is higher with 14 tools used. Tools that support multiple languages or frameworks like Dependabot, CodeQL, and Checkstyle, and Coverity are used more often than tools that focus on a specific use case or languages like Dart Analyzer, conftest, or Cookstyle. 
        According to the 2020 stackoverflow annually developer survey the top 5 used languages are JavaScript, HTML/CSS, SQL, Python, Java, C\#, Typescript, and PHP\cite{stackoverflow:2019}. All of the detected tools related to SA1, SA2, and SA3 support at least one of these languages.
        \begin{center}
            \fbox{\parbox[l]{0.9\columnwidth}{\begin{itemize}[leftmargin=*]
            \item Most used security tools require low integration effort and support multiple languages/frameworks.
            \item Each security tool supports at least one of the most popular languages.
            \end{itemize}
            }}
        \end{center} \vspace{0.2cm}
        
        \begin{table}[t]
            \caption{Distribution of Tool Usage per Security Activity}
            \begin{center}
                \begin{tabular}{l|c|c}
                    \hline
                    \textbf{Security} & \multicolumn{2}{|c}{\textbf{Security Tool Usage}} \\
                    \cline{2-3}
                    \textbf{Activity} & \textbf{Tool} & \textbf{Projects} \\
                    \hline
                    \hline
                    
                     & \textbf{Dependabot} & 169 \\
                    \textbf{3rd Party} & \textbf{PHP Security Checker} & 31 \\
                    \textbf{Vulnerability} & \textbf{Snyk} & 18 \\
                    \textbf{Scanning} & \textbf{Clair} & 11\\
                        & \textbf{Blackduck} & 8\\
                        & \textbf{Retire.JS} & 8\\
                        & \textbf{SRC:CLR} & 8 \\
                        & \textbf{Anchore} & 6 \\
                        & \textbf{Node Security Platform} & 5\\
                        & \textbf{Trivy} & 4 \\
                        & \textbf{Sonatype} & 3 \\
                        & \textbf{Whitesource} & 1 \\
                        \hline\hline
                        & \textbf{CodeQL} & 68 \\
                    \textbf{Static/Dynamic} & \textbf{CheckStyle} & 44 \\
                    \textbf{Application} & \textbf{Coverity} & 33 \\
                    \textbf{Security} & \textbf{SonarQube} & 27 \\
                    \textbf{Testing} & \textbf{Bandit} & 20 \\
                        & \textbf{PHPStan} & 19 \\
                        & \textbf{LGTM} & 8 \\
                        & \textbf{Brakeman} & 5 \\
                        & \textbf{Qark} & 2 \\
                        & \textbf{ESLint Security} & 1 \\
                        & \textbf{Dart Analyzer} & 1 \\
                        & \textbf{Checkmarx} & 1 \\
                        & \textbf{DevSkim} & 1 \\
                        & \textbf{CFN NAG} & 1 \\
                        \hline\hline
                        & \textbf{Chef} & 49 \\
                        \textbf{Secure} & \textbf{Ansible} & 36 \\
                        \textbf{Configuration/} & \textbf{Terraform} & 18 \\
                        \textbf{Hardening} & \textbf{Puppet} & 11 \\
                        & \textbf{Phan} & 15 \\
                        & \textbf{Seccomp} & 10 \\
                        & \textbf{CloudCustodian} & 5 \\
                        & \textbf{Turbot} & 1 \\
                        \hline\hline
                        & \textbf{Bane} & 15 \\
                        \textbf{Compliance/} & \textbf{Ansible-Lint} & 14 \\
                        \textbf{Hardening} & \textbf{Blackbox} & 12 \\
                        \textbf{Checks} & \textbf{Puppet-Lint} & 10 \\
                        & \textbf{InSpec} & 9 \\
                        & \textbf{Kube Audit} & 8 \\
                        & \textbf{Foodcritic} & 7 \\
                        & \textbf{Cookstyle} & 4 \\
                        & \textbf{conftest} & 3 \\
                        \hline\hline
                        & \textbf{Pinterest Knox} & 17 \\
                        \textbf{Secrets} & \textbf{GitLeaks} & 14 \\
                        \textbf{Management} & \textbf{CyberArk Conjur} & 14 \\
                        & \textbf{ThoughtWorks Talisman} & 12 \\
                        & \textbf{Berglas} & 10 \\
                        & \textbf{Docker Secrets} & 5 \\
                        & \textbf{Trufflehog} & 1 \\
                    \hline
                \end{tabular}
                \label{table:sec_tool_distr}
            \end{center}
        \end{table}

    \subsection{RQ3: Do certain types of projects perform security activities more often than others?}
\begin{figure}[b]
            \centerline{\includegraphics[width=\columnwidth]{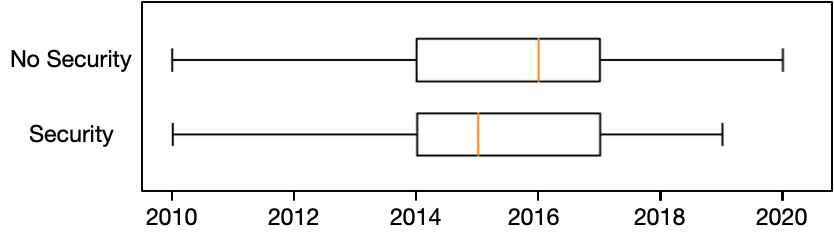}}
            \caption{Boxplot in which year the project was created comparing projects performing security to projects not performing security.}
            \label{fig:char_created}
        \end{figure}

\begin{figure}[b]
            \centerline{\includegraphics[width=\columnwidth]{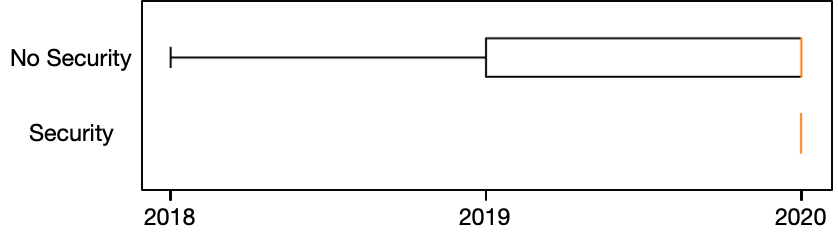}}
            \caption{Boxplot in which years the project was last updated comparing projects performing security to projects not performing security.}
            \label{fig:char_updated}
        \end{figure}

\begin{figure}[t]
            \centerline{\includegraphics[width=\columnwidth]{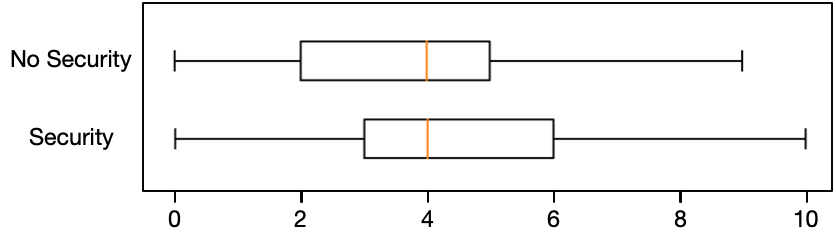}}
            \caption{Boxplot for how long the project is maintained in years comparing projects performing security to projects not performing security.}
            \label{fig:char_maintained}
        \end{figure}

        \noindent To analyze whether projects with certain characteristics perform security activities more often than others, we investigated: the number of stars of a project, when it was created, last updated, and for how long it is maintained. Additionally, we researched the languages used in the projects. The investigation of stars revealed that there are slight differences between projects using security tools and projects not using security tools. The lower quartile of the projects performing security activities had 11 stars more than the lower quartile of the projects not performing security activities. A similar, but smaller difference of 40.5 stars (with security) to 38 stars (no security) could be measured at the median. For both, projects with security and projects without security there is no difference in the upper quartile (327 stars). This leads to the conclusion, that the number of stars (as a measure of popularity) of a project seems to be rather unrelated to the usage of security tools. \Cref{fig:char_created} depicts the quartiles for the characteristic when the project was created. In \Cref{fig:char_updated} the distribution for the property when it was last updated is illustrated. The maintenance duration in years for each project is visualized in \Cref{fig:char_maintained}. Each of these characteristics is rounded to years, so an update in September 2017 is counted as an update in 2017. For the characteristic "created in year" the differences are negligible. However, of those projects performing security activities 78\% were last updated in 2020 while for those not performing security activities only 69\% were updated in 2020. Combined with the difference in number of years maintained we can deduce that there is a relation between projects performing security and how long they are maintained as well as how recently they have been updated.
        
\begin{figure}[b]
            \centerline{\includegraphics[width=0.88\columnwidth]{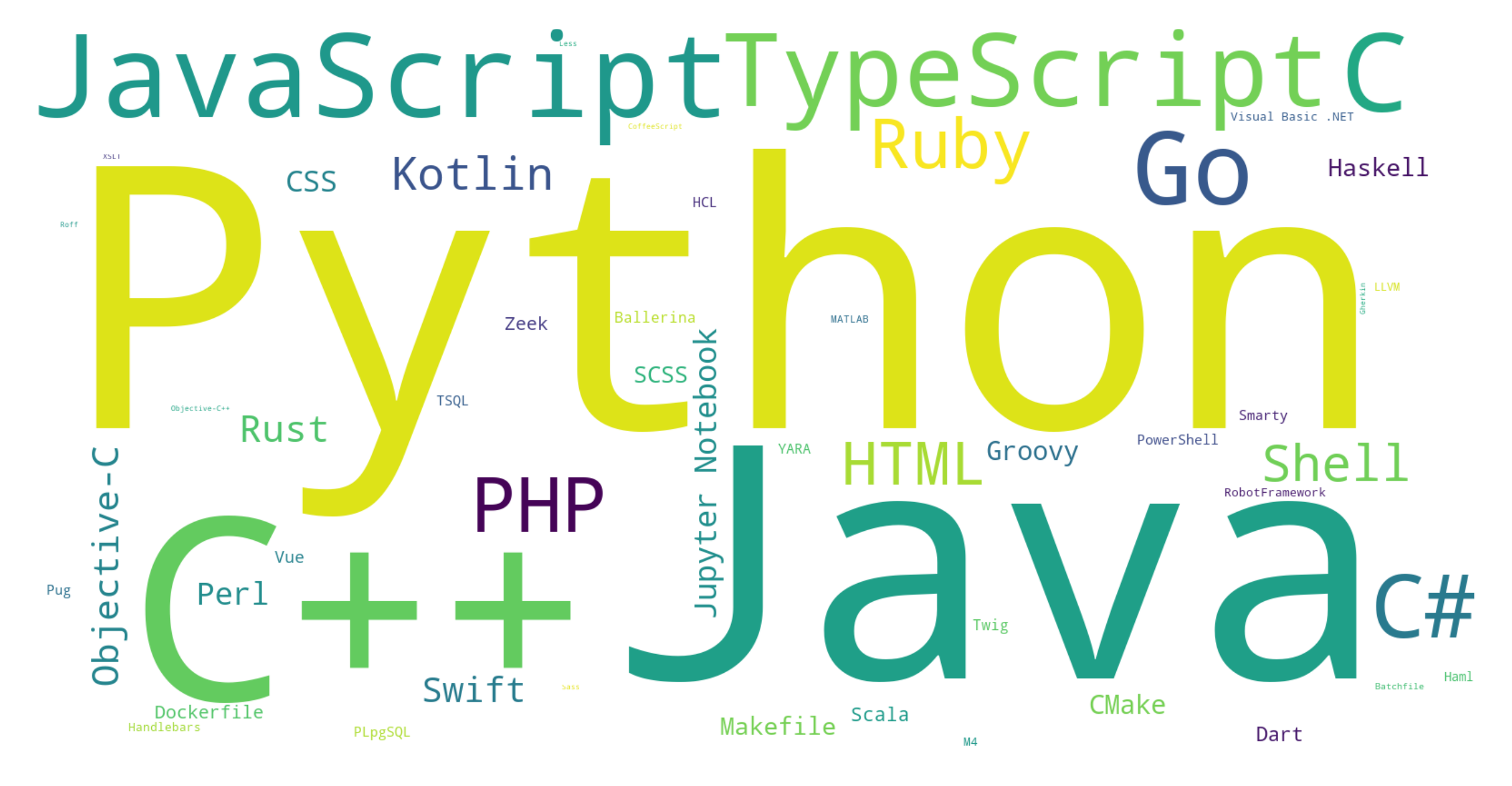}}
            \caption{Wordcloud of languages used in projects performing Static/Dynamic Application Security Testing.}
            \label{fig:wordcloud_sast_dast}
        \end{figure}

\begin{figure}[htbp]
            \centerline{\includegraphics[width=0.88\columnwidth]{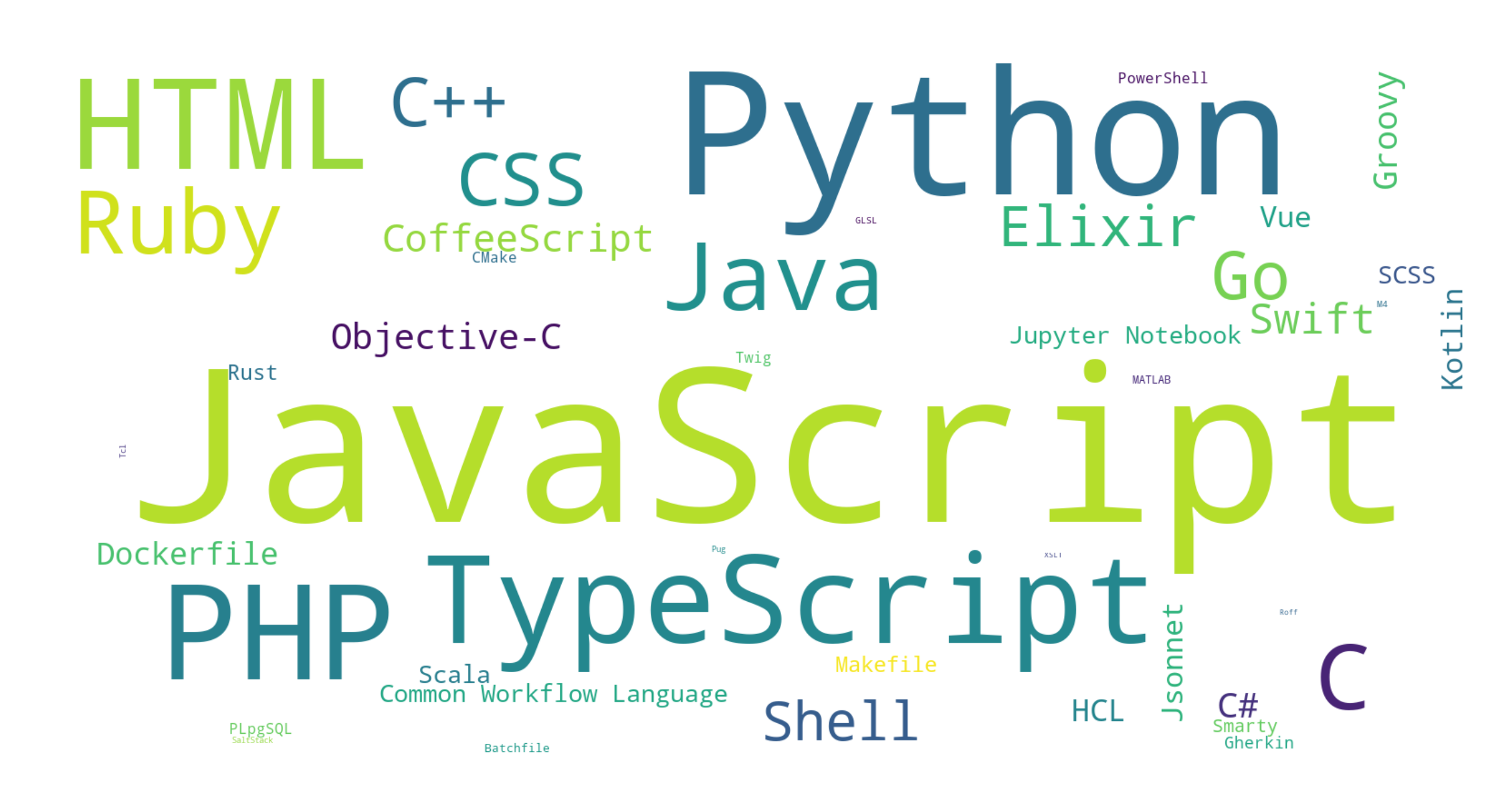}}
            \caption{Wordcloud of languages used in projects performing 3rd Party Vulnerability Scanning.}
            \label{fig:wordcloud_3rd_party}
        \end{figure}

\begin{figure}[htbp]
            \centerline{\includegraphics[width=0.88\columnwidth]{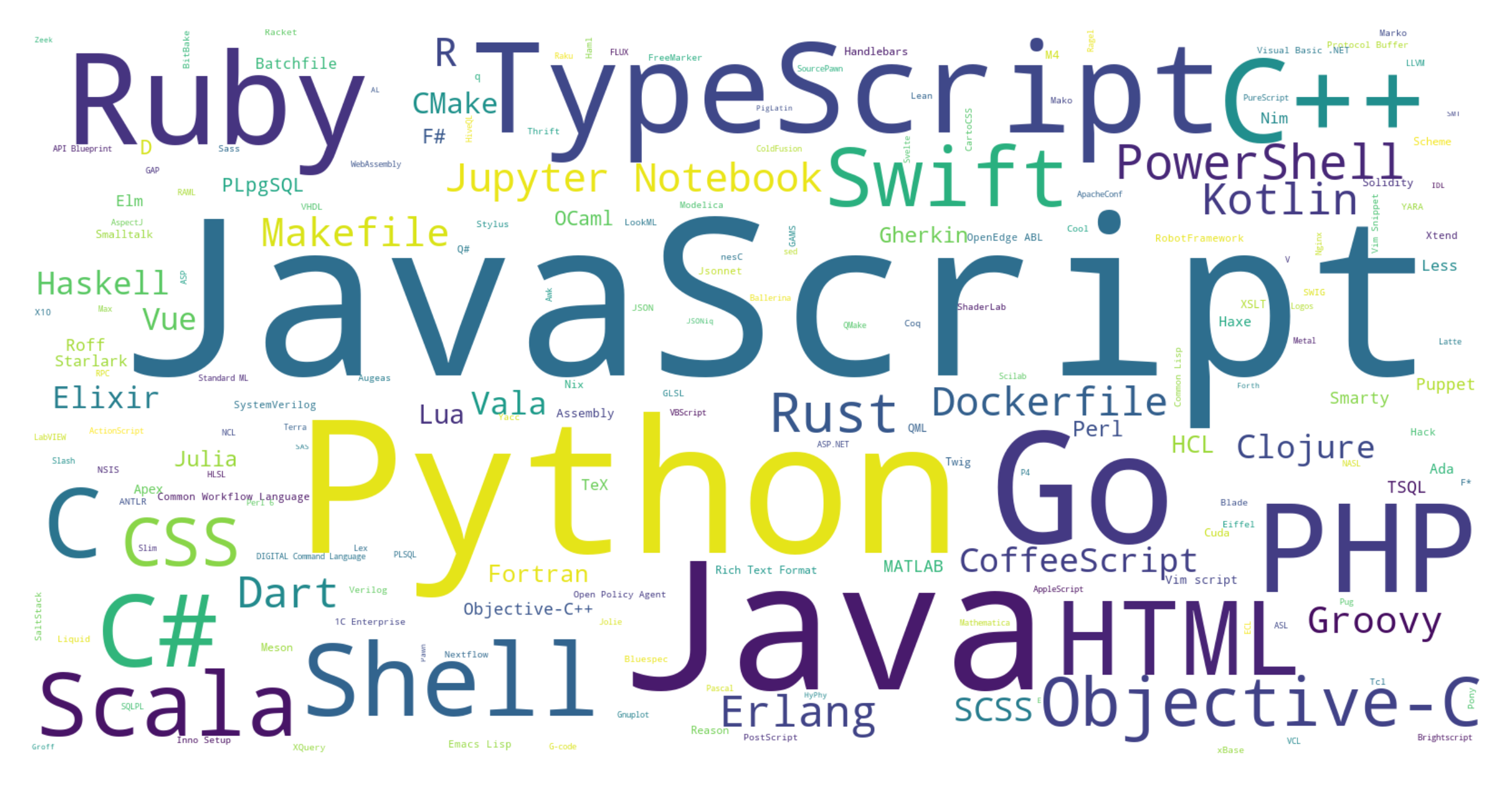}}
            \caption{Wordcloud of languages used in projects not performing Security.}
            \label{fig:wordcloud_no_sec}
        \end{figure}

        In the analysis of languages, we used wordclouds to visualize the differences between security activities. In a wordcloud the most frequent terms are highlighted with the biggest font, thus the bigger the language, the more often it is used. For simplicity we will only focus on the two most performed security activities here: 3rd Party Vulnerability Scanning and Static/Dynamic Application Security Testing. Each is depicted as a wordcloud, in \Cref{fig:wordcloud_3rd_party} and \Cref{fig:wordcloud_sast_dast} respectively. Additionally, we visualize the language usage of projects not performing security activities in \Cref{fig:wordcloud_no_sec}. When comparing the languages from Static/Dynamic Application Security Testing to those of 3rd Party Vulnerability Scanning the biggest differences are visible for Python, Java, C++, JavaScript, and TypeScript. While Static/Dynamic Application Security Testing is performed more often when programming in Python, Java, and C++ than in JavaScript or Typescript it is nearly the opposite when looking at 3rd Party Vulnerability Scanning with JavaScript, Python, TypeScript, and PHP being the dominant languages. One explanation could be that programming in TypeScript (thus also in JavaScript) contains extensive 3rd Party importing, compared to other programming languages. Nonetheless, these findings do not imply that given a project uses e.g. Java it performs Static/Dynamic Application Security Testing and no 3rd Party Vulnerability Scanning. That can also be seen in \Cref{fig:wordcloud_no_sec} where all languages used by projects not performing security are depicted. Here e.g. Java is also used often. Another insight is that languages like Ruby, Swift, and Scala are used a lot in projects not performing security activities, but not in projects performing security activities. One reason for this could be a lack of security tools. Unfortunately, we can not deduce the causes for the before mentioned findings of this research question. However, we set the baseline for future work on analyzing the reasoning on why projects using certain languages perform specific security activities more often.
        \begin{center}
            \fbox{\parbox[l]{0.9\columnwidth}{Programming languages have a direct relation to the types of security activities performed in a project.}}
        \end{center} \vspace{0.2cm}

\section{Discussion} \label{sec:discuss}
\subsection{Summary of conclusions}
\noindent This study evidences the state-of-practice of security in enterprise-driven open source software by means of security activities that are automated into continuous integration pipelines. Results provide visibility of which security activities are applied by providers during OSS development process (see \cref{table:sec_act_distr}). Industrial companies may use our results to adequately plan and further close the protection cycle for their software products. Moreover, DevOps and security practitioners can use the mapping of security activities and security tools (see \cref{table:sec_tool_distr}) for decision making of security automation tools to integrate into industrial CI pipelines. Although several other aspects should be considered to justify such selection, the study results serve practitioners as precise evidence, especially in regulated environments, where security standards demand concrete explanations about tailoring of secure development processes. In the following, we discuss the study results and how they relate with our observations as security practitioners who focus on security compliance in industrial agile and DevOps projects.

\textbf{Automated security activities in enterprise-driven OSS projects are scarce and protection coverage is rather low}. While 6.83\% of 8,423 analyzed projects perform security activities, only 1\% of these projects perform more than 2 security activities. This alerts industry practitioners to keep a strong focus on detecting vulnerabilities in 3rd Party software. 

\textbf{3rd Party Vulnerability Scanning (SA1) and Static/Dynamic Application Security Testing (SA2) are the top security activities} in enterprise-driven OSS software performed by 3\% and 2.7\% of the analyzed projects. With regard to SA1, this makes explicit that even OSS providers are concerned about security issues on 3rd Party software. We know that trusting 3rd-Parties is indeed a top concern in industry, therefore we infer it is similar in OSS projects. However, we discuss the incomplete approach of only using tools to solve the problem. Although tools help to detect vulnerabilities, the protection cycle closes only when vulnerabilities are solved by providers. Practitioners should establish processes to get on-time notification of security issues from OSS providers, as well, as to accurately delivering patches to clients. Regarding SA1 tools, OSS projects integrate some security tools that we also observe in industry, although not with the same order of preference. Referring to languages, a main difference is that industry applies SA1 extensively, not only for JavaScript and Python but also for languages like Java, C, C++, C\# (see \Cref{fig:wordcloud_3rd_party}).
Same as for OSS projects, SA2 is frequent in CI industrial pipelines. First, finding bugs is a mature practice that includes finding application vulnerabilities. Secondly, regulators explicitly demand SA2 e.g. static code analysis as application testing in security standards like IEC 62443-4-1 \cite{iec4_1}. Contrary to the evidence for the analyzed OSS projects, industry uses frequently more specialized and licensed SA2 tools. Language coverage is also broader in industry, although covering the ones mentioned in the analyzed projects \Cref{fig:wordcloud_sast_dast}.

\textbf{Secure Configuration/Hardening (SA3) and Compliance Checks (SA4) are not well covered }in enterprise-driven OSS, performed by 1.3\% and 0.6\% of the analyzed OSS projects. For both activities, OSS practitioners refer that such activities pertain to the user of the OSS component. OSS component users should harden the intended environment for the component themselves. In addition, SA4 may not even be important for OSS providers as they do not deal with regulators, who demand compliance with best practices. Industry practice is more mature in this matter, environments are continuously hardened based on best practices, both in cloud and on-premise. However, industry sometimes faces the issue that OSS components do not run in well-hardened environments. Therefore, OSS providers, serving regulated environments, should also implement SA3 and SA4, as part of the industrial product lifecycle. Regarding SA3 and SA4 tools, we observe frameworks that allow for secure configuration even though they are not considered security tools per see.

\textbf{SA5 Secrets management is a missing security activity}, performed by 0.4\% of the studied projects. This provides industry security practitioners, who analyze threat models, a more clear notion about the risk that an OSS provider credential is compromised. Besides, they may validate the probability of other typified attack vectors in industry such as exploitation of credential access or forced authentication \cite{mitre:2020}. Contrasting SA5 with SA1 results, OSS developers seem to focus more on security that 3rd-Parties offer, rather than focusing on achieving basic security steps like adequately managing credentials or passwords. Besides security activities automation, we observe that OSS projects mainly use only two of the continuous integration services that we observe in industrial DevOps projects (see \cref{table:CI_service_usage}). Also, our study results differ from Widder's et al. \cite{8595199}, who refer that Travis CI is abandoned in OSS projects.

\noindent In summary, we observe that enterprise-driven OSS projects need to extend the integration of security activities in their CI pipelines. Based on repositories analysis and answers from professionals, we conclude not to focus only on the integration of security tools, but also on training development professionals. Industrial companies, that utilize OSS components, may consider complementing their security strategies by increasing awareness of OSS providers with regard to best practices.

\subsection{Threats to Validity}
\noindent As in any empirical study, the one at hands face validity threats. In this section, we present the most relevant ones.
The main threat refers to the use of an existing data set, which was not specifically gathered for this study. To prove validity, we use a research data set that is the basis of a peer-reviewed scientific publication \cite{Spinellis:2020}. Later, we report results based on three layers of analysis: crawling, manual review, and direct contact with maintainers. \\
To address researcher bias, two authors performed data collection and results mapping while the two most experienced authors interpreted and concluded based on industry experience. 
We avoid conflicts of interest with security tool providers by remaining neutral and not discussing the pros or cons of any specific security tool. Finally, we followed best practices for empirical studies \cite{ralph:2020,Wagner2019ChallengesIS,ben2017empirical} and study rigor is controlled by an experienced empirical methods researcher. \\
The main limitation of this study is its focus on open source projects without insights on close source projects. We did not explore the OSS projects to determine if their components are being used in existent industrial products. Also, we are aware of possible improvements to automated crawling and detection scripts.

\section{Conclusion and Future Work}
    \label{sec:conclusion} 
\noindent This paper provides evidence on security activities integration in enterprise-driven open source software projects. We mine a validated dataset (c.f. Spinellis et al.\cite{Spinellis:2020}) rendering the state-of-practice based on automatic analysis of 8,243 projects, with manual validation of 400 projects and survey-based confirmation of 67 projects. Further, we discussed implications for industrial companies.\\
The main results were discussed in the previous section, and can be summarized as follows:
\begin{enumerate}
 \setlength\itemsep{2pt}
    \item even though OSS project maintainers perceive security as rather important only 6.83\% of them perform security activities in continuous integration. This means security issues may be overseen in the development process of OSS.
    \item reasons for not performing security are that maintainers see responsibility for security on side of the user/integrator and that the attack surface is perceived as too limited.
    \item integration of security activities may be influenced by programming languages and training of maintainers.
\end{enumerate}

\noindent Our contribution depicts also the most used security tools in OSS projects, which serve practitioners on further analysis. To allow researchers to replicate or extend the study, our artefacts are publicly disclosed and available \cite{we:2020}.\\

\noindent As future work, we will use empirical methods to analyze the top used tools in OSS projects per activity. We hypothesize that security tools that are mostly used, are more mature, the integration is simpler, or even they cover a first security priority. Later, we can compare them with tools that are already used in industrial projects. This will improve our ability to recommend security tools to practitioners.

\noindent Besides, we aim to explore the suitability to use security tools artefacts as evidence of compliance of security activities as explain by standards. This will complement our research on security compliant agile and DevOps (c.f.\cite{moyon2020compliantagile,moyon2020compliantdevops}). 
 
 \section*{Data Availability}
 \noindent All artifacts are made available online and can be accessed at: \url{https://doi.org/10.5281/zenodo.4106329} \cite{we:2020}.

\section*{Acknowledgment}
\noindent The authors would like to thank the project maintainers that participated in the survey. Moreover, the authors acknowledge industry security practitioners who endorse to empirical studies concerning secure compliant agile and DevOps. \\
Parts of this work was supported by the KKS foundation through the S.E.R.T. Research Profile project at Blekinge Institute of Technology and SERL Lab.

\bibliography{main}
\bibliographystyle{IEEEtran}

\end{document}